\documentclass[a4paper,11pt,fleqn]{article}
\usepackage[official]{eurosym}
\usepackage{amsmath}
\usepackage{amssymb}
\usepackage{theorem}
\usepackage[usenames,dvipsnames]{pstricks}
\usepackage{euscript}
\usepackage{graphicx}
\usepackage{charter}

\usepackage{comment}
\usepackage{pgfplots,upgreek,subfigure}

\newcommand{\email}{}
\definecolor{labelkey}{rgb}{0,0.08,0.45}
\definecolor{refkey}{rgb}{0,0.6,0.0}
\definecolor{Brown}{rgb}{0.45,0.0,0.05}
\definecolor{dgreen}{rgb}{0.00,0.49,0.00}
\definecolor{dblue}{rgb}{0,0.08,0.75}
\RequirePackage[colorlinks,hyperindex]{hyperref} 
\hypersetup{linktocpage=true,citecolor=dblue,linkcolor=dgreen}
\usepackage{cleveref}
\PassOptionsToPackage{normalem}{ulem}
\usepackage{ulem}

\renewcommand{\geq}{\ensuremath{\geqslant}}
\renewcommand{\le}{\ensuremath{\leqslant}}
\renewcommand{\ge}{\ensuremath{\geqslant}}

\theoremstyle{plain}{\theorembodyfont{\rmfamily}%
}
\theoremstyle{plain}{\theorembodyfont{\rmfamily}%
}
\theoremstyle{plain}{\theorembodyfont{\rmfamily}%
}
\theoremstyle{plain}{\theorembodyfont{\rmfamily}%
}
\theoremstyle{plain}{\theorembodyfont{\rmfamily}%
}
\theoremstyle{plain}{\theorembodyfont{\rmfamily}%
}
\theoremstyle{plain}{\theorembodyfont{\rmfamily}%
}
\theoremstyle{plain}{\theorembodyfont{\rmfamily}%
}

\numberwithin{equation}{section}

\colorlet{color1}{cyan!20}
\colorlet{color2}{yellow!20}

\colorlet{color1dark}{cyan!50!black}
\colorlet{color2dark}{yellow!50!black}

\colorlet{colorreject}{green!20}
\definecolor{blue1}{rgb}{0.004, 0.451, 0.698}
\definecolor{brown1}{rgb}{0.835, 0.369, 0.0}
\definecolor{brown2}{rgb}{0.008, 0.620, 0.451}
\colorlet{colorcannot}{red!20}
\definecolor{green1}{rgb}{0.871, 0.561, 0.0196}

\colorlet{colorrejectdark}{green!50!black}
\colorlet{colorcannotdark}{red!50!black}

\newcommand{\xmap}{\pmb{x}^{\dagger}}
\newcommand{\fwd}{\pmb{\upphi}}
\newcommand{\reg}{\pmb{\uppsi}}
\newcommand{\noise}{\pmb{w}}
\newcommand{\data}{\pmb{y}}
\newcommand{\im}{\pmb{x}}

\newcommand{\xs}{\pmb{x}_{\mathcal{S}}}
\newcommand{\xc}{\pmb{x}_{\mathcal{C}}}
\newcommand{\xsh}{\hat{\pmb{x}}_{\mathcal{S}}}
\newcommand{\xch}{\hat{\pmb{x}}_{\mathcal{C}}}

\newcommand{\estimate}{\hat{\theta}}
\newcommand{\Calpha}{\mathcal{C}_{\alpha}}
\newcommand{\Sset}{\mathcal{S}}
\newcommand{\hpd}{\eta_{\alpha}}
\newcommand{\mask}{\textbf{M}}
\newcommand{\radgrad}{r_{\nabla}}
\newcommand{\radpix}{r_{\text{pix}}}
\newcommand{\mugrad}{\mu_{\nabla}}
\newcommand{\mupix}{\mu_{\text{pix}}}
\newcommand{\xmapS}{\pmb{x}^{\dagger}_{\mathcal{S}}}
\newcommand{\rhoal}{\rho_{\alpha}}
\newcommand{\Ndct}{D}
\newcommand{\Nangle}{M_a}

\pgfmathdeclarefunction{gauss}{2}{%
  \pgfmathparse{1/(#2*sqrt(2*pi))*exp(-((x-#1)^2)/(2*#2^2))}%
}

\begin{document}


\title{\sffamily Uncertainty Quantification in CT pulmonary angiography}
\author{
Adwaye M Rambojun$^\star$,
Hend Komber$^\circ$,
Jennifer Rossdale$^\circ$,
Jay Suntharalingam$^{\circ\bullet}$, \\
Jonathan C L Rodrigues$^\circ$,
Matthias J Ehrhardt$^\star$, and
Audrey Repetti$^{\dagger\ddagger}$\footnote{M.J.E. and A.R. are joint senior authors.}
\\[5mm]
\small
\small $^\star$ Mathematical Sciences, University of Bath, UK\\
\small $^\circ$ Royal United Hospital, Bath, UK \\
\small $^\bullet$ Life Sciences, University of Bath, UK \\
\small $^\dagger$ Institute of Sensors, Signals and Systems, Heriot-Watt University, Edinburgh, UK\\
\small $^\ddagger$ Maxwell Institute for Mathematical Sciences, Edinburgh, UK\\
\small 
E-mail: \email{a.repetti@hw.ac.uk}, \email{me549@bath.ac.uk}
}
\date{}

\maketitle

\vskip 8mm

\begin{abstract}
Computed tomography (CT) imaging of the thorax is widely used for the detection and monitoring of pulmonary embolism (PE). 
However, CT images can contain artifacts due to the acquisition or the processes involved in image reconstruction. 
Radiologists often have to distinguish between such artifacts and actual PEs. 
Our main contribution comes in the form of a scalable hypothesis testing method for CT, to enable quantifying uncertainty of possible PEs. 
In particular, we introduce a Bayesian Framework to quantify the uncertainty of an observed compact structure that can be identified as a PE. 
We assess the ability of the method to operate under high noise environments and with insufficient data. 
\end{abstract}

{\bfseries Keywords.}
Medical Imaging $|$ Hypothesis Testing $|$ Optimization $|$ Bayesian $|$ Pulmonary Embolism



\subsection*{Significance statement}
Computed Tomography (CT) imaging in medicine is widely used to visualize internal organs for diagnostic purposes. In the context of pulmonary embolism (PE) detection in the setting of acute chest pain, the PE can appear in CT scans as small structures with weak amplitude. 
So PE detection can be challenging in practice for clinicians, who have to decide whether the structures are PEs or not. This ambiguity can occur due to imperfect data acquisition (e.g. insufficient data, high noise environment). 
In this work we propose a computational tool to help clinicians to decide whether an observed structure is a PE or an artifact due to imperfect data. Our method quantifies the uncertainty of the structure, leveraging optimization and Bayesian theory.

\section{Introduction}

\subsection{PE detection with computed tomography angiography} 

Most medical image modalities such as Computed Tomography (CT), ultrasound and Magnetic Resonance Imaging are the result of an intricate image reconstruction process that uses noisy and incomplete captured data.
In particular, CT is a popular imaging modality used to diagnose various types of pathologies, such as acute inflammatory conditions, strokes and malignancy. X-rays are passed through the patient's body from multiple angles and an attenuation coefficient is calculated depending on the densities of the different tissues the x-rays pass through. A reconstruction algorithm is used to create final 3D image. This algorithm is subject to creation of artifacts, i.e., structures not present in the ground truth image being captured \cite{withers2021x}. 
They can interfere with conclusions drawn by radiologists, who then have to infer if structures appearing in CT images are pathological or artifactual due to the inaccuracy of the data acquisition.


This is quite common when assessing CT scans for the presence of acute PE, which is a major cause of mortality with  approximately 30,000 deaths per year in the UK \cite{house2005report}. Assessment and detection of PE and its cardiovascular complications is routinely performed with a CT pulmonary angiography (CTPA)~\cite{meinel2015predictive}. Chronic thromboembolic pulmonary hypertension is also a potential long term disabling complication of acute PE and CTPA is an important diagnostic tool as well as being useful to assess for operability~\cite{Kim1801915}. However, a variety of patient and protocol related factors can result in image artifacts that may impact the clinicoradiological confidence of image interpretation. If a false positive diagnosis is made, this can result in inappropriate patient treatment with anticoagulation, which is associated with an unnecessary increase in bleeding rates \cite{kempny2019incidence}. 

In this context, quantifying uncertainty of the PE-like structures observed in reconstructed CT thorax images would improve diagnosis accuracy. 
In this paper we present an uncertainty quantification (UQ) framework to perform hypothesis tests on PE-like structures, and determine whether they are present in the patient thorax or are artifacts arising from inaccurate data acquisition. 

\subsection{Bayesian inference for imaging}
Reconstruction of images from CT data can be formulated as an inverse problem. The objective is to find an estimate $\xmap$ of an unknown image~$\im$ (i.e., patient's thorax) from measurements~$\data$ acquired with a CT scanner \cite{Seeram2015, Hansen2021}. 
Following a Bayesian framework~\cite{Kaipio2006}, the image and the data are related through a statistical model. Then the estimate $\xmap$ is inferred from $\data$ according to its posterior distribution, which combines information from the likelihood, related to the observations $\data$, and the prior, used to introduce \textit{a priori} information on the target image. 
The prior is used to regularize the model, to help to overcome ill-posedness and/or ill-conditionedness of the inverse problem. Common choices are to impose feasibility constraints, and to promote smoothness or sparsity of $\im$, possibly in some transformed domain such as wavelet, Fourier or total variation (TV) \cite{Bredies2018book}.

Sampling methods, e.g., Markov Chain Monte Carlo methods (MCMC), draw random samples according to the posterior distribution. These methods then allow us to form estimators (e.g., minimum mean square error (MMSE) estimator, posterior mean or maximum a posteriori (MAP) estimator), and to perform UQ through confidence intervals and hypothesis testing \cite{robert2007bayesian, robert2004monte}.  
The main drawback of these methods is their high computational cost making them inefficient for high-dimensional problems, as encountered in imaging. Indeed, for CT imaging, the dimension of $\im$ are often of the order of $10^8$ in the case of high resolution lung scans \cite{siemensManual}. Although multiple works have emerged in the last years to help scaling sampling methods, e.g. \cite{pereyra2016proximal, vono2020, thouvenin2022}, they usually remain prohibitive in such high dimensions.

Methods of choice for handling high-dimensional problems are proximal splitting optimization algorithms \cite{combettes2011proximal, komodakis2015playing, Chambolle2016actanumerica}. These are known to be very efficient to form MAP estimates. Nevertheless, these methods only provide a point estimate, without quantifying the uncertainty on the delivered solution. To overcome this issue, recently a Bayesian Uncertainty Quantification by Optimization (BUQO) approach has been proposed in \cite{pereyra2017maximum, repetti2018eusipco, repetti2019scalable}, to perform hypothesis testing on particular structures appearing on MAP estimates. The method determines whether the structures of interest are true, or are reconstruction artifacts due to acquisition inaccuracy. BUQO has the advantage of being scalable for high-dimensional problems, as the UQ problem is recast in an optimization framework, to leverage
proximal splitting optimization algorithms. 

\subsection{Uncertainty Quantification for PE}
UQ is the main tool to assist doctors for accurate decision-making processes. Ill-posed and ill-conditioned inverse problems result in high uncertainty about the estimate. 
In this work, we focus on quantifying uncertainty of PE-like structures in CT thorax images. Specifically, we design a method based on BUQO to determine whether these structures are PEs, or if they are reconstruction artifacts.

\section{Methods}
\label{Sec:method}

In this section we describe the steps of the proposed PE UQ technique.
First, we form the CT image using an optimization algorithm (Section~\ref{Sec:method}\ref{Ssec:method:CTimaging}). 
Second, we identify PE-like structures in the image estimate, and postulate the null hypothesis that these structures are not present in the ground truth image, i.e., they are not in the patient's thorax, but instead are reconstruction artifacts arising due to the ill-posedness of the problem. Third, we use our method to decide whether the null-hypothesis can be rejected or not (Section~\ref{Sec:method}\ref{Ssec:method:hyp}).

\subsection{Bayesian inference and optimization for CT imaging}
\label{Ssec:method:CTimaging}

In general, the gantry of a CT scanner, which includes multiple x-ray sources and multiple detectors will rotate around the patient's chest. This generates an $M$-dimensional array of data, denoted by $\data$, consisting of attenuated X-ray intensities \cite{Seeram2015, Hansen2021}. The pattern of attenuation is determined by the geometry of the area through which the beams are directed. The aim of CT reconstruction is to recover a voxel array of dimension $N$\footnote{Here $N$ is the product of the individual dimensions of the 3D voxel array.}, denoted by $\im$, that represents the geometry of the organs inside the thorax given the observed noisy data $\data$. 
This can be reasonably approximated as a linear inverse problem of the form
\begin{equation}    \label{pb:inv}
\pmb{y} = \fwd \im + \noise
\end{equation}
where $\fwd$ represents the CT measurement operator described above, and $\noise$ is a realization of an additive independent and identically distributed (i.i.d.) random noise. 

Using a Bayesian formulation, the posterior distribution of the problem, which combines information from the likelihood and the prior, can be expressed as
\begin{equation}\label{eq-bayes}
p(\pmb{x} | \data) \propto \exp( -f_{\data}(\fwd\pmb{x}) - g(\pmb{x})),
\end{equation}
where $f$ is assumed to be a log-concave likelihood associated with the statistical model of~\eqref{pb:inv}, and $g$ is a log-concave prior distribution for $\im$. 
The usual approach to estimate $\im$ is to use a MAP approach, that consists in defining $\xmap$ as a minimizer of the negative logarithm of~\eqref{eq-bayes}, i.e.,
\begin{equation}    \label{pb:mingen0}
    \xmap \in \underset{\pmb{x}}{\mathrm{argmin}} \;\; f_{\data}( \fwd \pmb{x}) + g(\pmb{x}).
\end{equation}

In this work, we assume that the exact noise distribution is unknown, but that it has a bounded energy, i.e., $\| \noise \|_2 \le \varepsilon$, where $\|\cdot\|_2$ is the usual Euclidean norm, and $\varepsilon>0$. Then, a typical choice is to take $f_{\data}( \fwd\pmb{x}) $ to be the indicator function of the $\ell_2$-ball $\mathcal{B}_2(\data, \varepsilon)$, centered in $ \data$ with radius $\varepsilon>0$. In addition, a common choice for the prior term $g(\pmb{x})$ is to promote sparsity of the image of interest in some basis (e.g., wavelet or TV). Then, \eqref{pb:mingen0} can be rewritten as
\begin{equation} \label{pb:mingen}
    \text{find } \xmap = \underset{\pmb{x}}{\mathrm{argmin}} \; \;  \| \reg \pmb{x}\|_1 \text{ s.t. } \| \fwd \pmb{x} - \data \|_2 \le \varepsilon,
\end{equation}%
where 
the operator $\reg$ models a linear transform, chosen such that $\reg \im$ has only few non-zero coefficients.
\eqref{pb:mingen} can be solved efficiently using proximal splitting algorithms~\cite{combettes2011proximal, komodakis2015playing, Chambolle2016actanumerica}.

\begin{figure}[t]
\begin{center}
\includegraphics[width=0.5\textwidth]{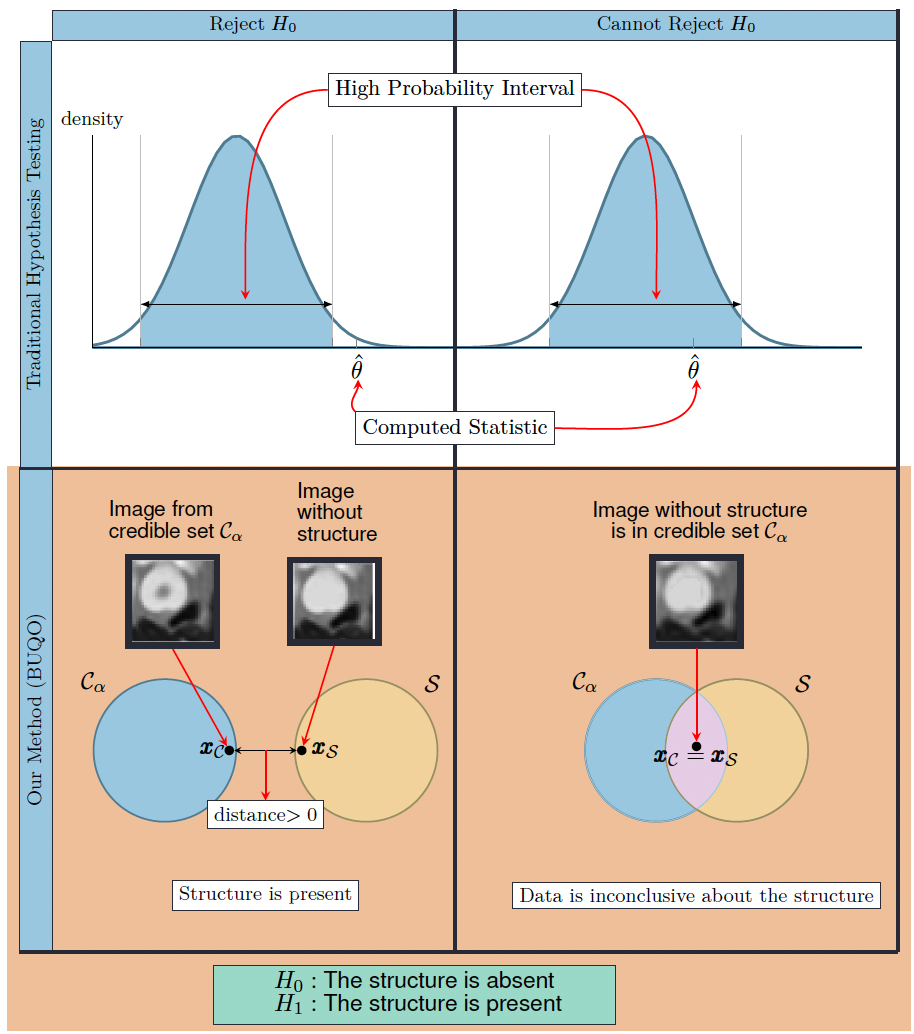}
\end{center}
\caption{The Figure shows the parallel between traditional hypothesis testing and our method. In traditional hypothesis testing, one computes the credible interval $\mathcal{I}_{\alpha}$ and the test statistic $\estimate$ from data. The null hypothesis $H_0$ is rejected if $\estimate$ is not in the credible region. Similarly for our proposed method, we compute the High Posterior Density (HPD) region $\Calpha$ and an image $\xs$ with the structure removed (similar to the test statistic). We reject the null hypothesis (which states that the structure is absent) if $\xs$ does not lie inside the credible region. This is determined by the distance between $\xs$ and $\xc$, which are the two elements of $\Sset$ and $\Calpha$ respectively that are closest to each other. If this distance is zero, we conclude that $\xc \in \Sset$ otherwise, $\xc \notin \Sset$.}
\label{fig-comparison}
\end{figure}

\subsection{High dimensional hypothesis testing}
\label{Ssec:method:hyp-PE}

The method described in the previous section provides a point estimate $\xmap$ of $\im$, without additional information regarding its uncertainty. 
In this work, we propose to perform a hypothesis test on structures that can be identified as PEs in the MAP estimate.

To illustrate our approach, we recall the basics of hypothesis testing. 
Typically, we postulate a null-hypothesis, i.e., we make a claim about the distribution of observed data. We use the observed data to compute a statistic $\estimate$. We decide to reject or not the null-hypothesis depending if $\estimate$ lies in a High Probability Interval (see Figure~\ref{fig-comparison}).

This can be extended to computational imaging~\cite{repetti2018eusipco, repetti2019scalable}, to quantify uncertainty on structures appearing on the MAP estimate $\xmap$, obtained by solving~\eqref{pb:mingen}. 
%
In this context, we postulate the null-hypothesis $\text{H}_0$ and the alternative hypothesis $\text{H}_1$ as follows:

\noindent
$\text{H}_0$: \textit{The structure is \textsc{absent} from the true image} \\ 
$\text{H}_1$: \textit{The structure is \textsc{present} in the true image}

\noindent
Formally, using Bayesian decision theory \cite{robert2007bayesian}, we can conclude that $\text{H}_0$ is rejected in favor of $\text{H}_1$ if $\mathbb{P}(\text{H}_0 | \data) \le \alpha$, where $\alpha \in (0,1)$ denotes the level of significance of the test. Such probability can be approximated by MCMC approaches \cite{robert2004monte}, however it becomes intractable for high-dimensional problems such as CT imaging.
To overcome this difficulty, we introduce a subset $\Sset$ of $\mathbb{R}^N$, associated with $\text{H}_0$, containing all the possible images without the structure of interest. 
Then, by definition, we have $\mathbb{P}(\text{H}_0 | \data) = \mathbb{P}( \im \in \Sset  | \data)$. 
To perform the hypothesis test, we will compare $\Sset$ with a posterior credible set $\Calpha^*$, corresponding to the set of possible solutions where most of the posterior probability mass of $\pmb{x}|\data$ lies \cite{pereyra2017maximum}.  
Formally, $\Calpha^*$ satisfies $\mathbb{P}(\im \in \Calpha^* |\data) = 1-\alpha$. Again computing such probability in high-dimension is intractable. Instead, \cite{pereyra2017maximum} introduced a conservative credible region $\Calpha$, in the sense that $\mathbb{P}(\im \in \Calpha |\data) \ge 1-\alpha$, that does not require any additional computational cost other than building a MAP estimate $\xmap$, i.e., solving~\eqref{pb:mingen}. Note that, by construction, we have $\xmap \in \Calpha$, and $\Calpha$ consists of defining a feasibility set around $\xmap$.


The BUQO approach adopted in this work consists in determining if the intersection between $\Sset$ and $\Calpha$ is empty. If it is empty, it means that $ \mathbb{P}( \im \in \Sset  | \data) = \mathbb{P}(\text{H}_0 | \data)  \le 1 - (1-\alpha) = \alpha$, hence $\text{H}_0$ is rejected. 
To determine if $\Sset \cap \Calpha = \varnothing$, we aim to find an image belonging to $\Sset \cap \Calpha$. If such image exists, it means that $\Sset \cap \Calpha \neq \varnothing$, and it is possible to find (at least) one image supported by the data $\data$ without the structure of interest, hence $\text{H}_0$ cannot be rejected. Otherwise $\Sset \cap \Calpha = \varnothing$, and $\text{H}_0$ is rejected (see the second row of Figure~\ref{fig-comparison}).

%

\subsection{Hypothesis test for PE detection}
\label{Ssec:method:hyp}

In this section, we explain the proposed method to determine whether $\Sset \cap \Calpha$ is empty or not. In addition, we give mathematical definitions of sets $\Sset$ and $\Calpha$, tailored for the PE UQ problem.

To find the closest image to the the MAP estimate $\xmap$, belonging to $\Sset$, one can project $\xmap$ into $\Sset$. We denote $\xmapS = \text{Proj}_{\Sset}(\xmap)$ this image. The first step is to verify if $\xmapS \in \Calpha$. If it is the case, then we have found an image in the intersection $\xmapS \in \Sset \cap \Calpha$, and $\text{H}_0$ cannot be rejected, i.e., we are uncertain that the PE is present. 
If $\xmapS \not\in \Calpha$, it does not mean that $\Calpha \cap \Sset$ is empty, and there might still be an image which belongs to both sets. To ascertain if the intersection is empty, we propose to equivalently compute the distance between $\Sset$ and $\Calpha$, denoted $\text{dist}(\Sset, \Calpha)$, and to verify if it is zero or positive. 
If $\text{dist}(\Sset, \Calpha)>0$, then we can conclude that $\Calpha \cap \Sset = \varnothing$, so $\text{H}_0$ is rejected in favor of $\text{H}_1$. Otherwise, if $\text{dist}(\Sset, \Calpha)=0$, there exists (at least) one image in the intersection, and hence $\text{H}_0$ cannot be rejected. 

To evaluate $\text{dist}(\Sset, \Calpha)$, we need to minimize the distance between an element $\xc$ of $\Calpha$ and an element $\xs$ of $\Sset$, i.e., we want to 
\begin{equation}
\text{find } (\xsh,\xch) = 
\underset{\xs \in \Sset, \, \xc \in \Calpha}{\mathrm{argmin}} \; \; 
\frac12 \| \xc - \xs \|_2^2 .
\end{equation}
For our problem, the conservative credible set, associated with~\eqref{pb:mingen}, is defined as $\Calpha :=\{ \im\ge 0  \,| \, \| \fwd \im - \data\|_2 \le \varepsilon \text{ and } \| \reg\im\|_1 \le \hpd   \}$, where
$\hpd = \|\reg \xmap\|_1 + N + \sqrt{16N\log \left(3/\alpha\right)}$. 
One main contribution of this work is to define $\Sset$ to be a set describing all possible images without PE structures that can be identified in the MAP estimate.
In particular, we want the pixel intensity profile within the structure's area to be similar to the pixel intensity profile of a neighborhood of the structure. To this aim, we propose to define $\Sset$ as the intersection of three sets, i.e., $\Sset := I \cap E \cap S$, given by 
\begin{align}
&\text{intensity:} & I &:=  \{\im \mid \im \geq 0\},    \label{def:int}\\
&\text{energy:} & E &:=  \{\im \mid \|\mask \im -\mupix \|_2< \radpix\},  \label{def:energy}\\
&\text{smoothness:} & S &:=  \{\im \mid \| \mask \nabla \im -\mugrad\|_2 < \radgrad\},  \label{def:smooth}
\end{align}
where $\mask \colon \mathbb{R}^N \to \mathbb{R}^{N_S}$ is a linear operator selecting the pixels of the image corresponding to the PE area. 
The first set $I$ is the positive orthant, to ensure images in $\Sset$ are intensity images. 
The second set $E$ controls the energy in the structure, ensuring that pixels inside the structure's area are taking values around a predefined mean value $\mupix$, chosen according to its neighborhood. 
The third set $S$ is a smoothness constraint, to control the pixel intensity variation in the structure's area to be close to a mean value $\mugrad$ corresponding to the variations in its neighborhood. For both $E$ and $S$, $\radpix$ and $\radgrad$ are positive predefined constants to control the similarity between the structure's area and its neighborhood.

\section{Experiments}

\begin{figure*}[t!]
\begin{center}
\includegraphics[width=\textwidth]{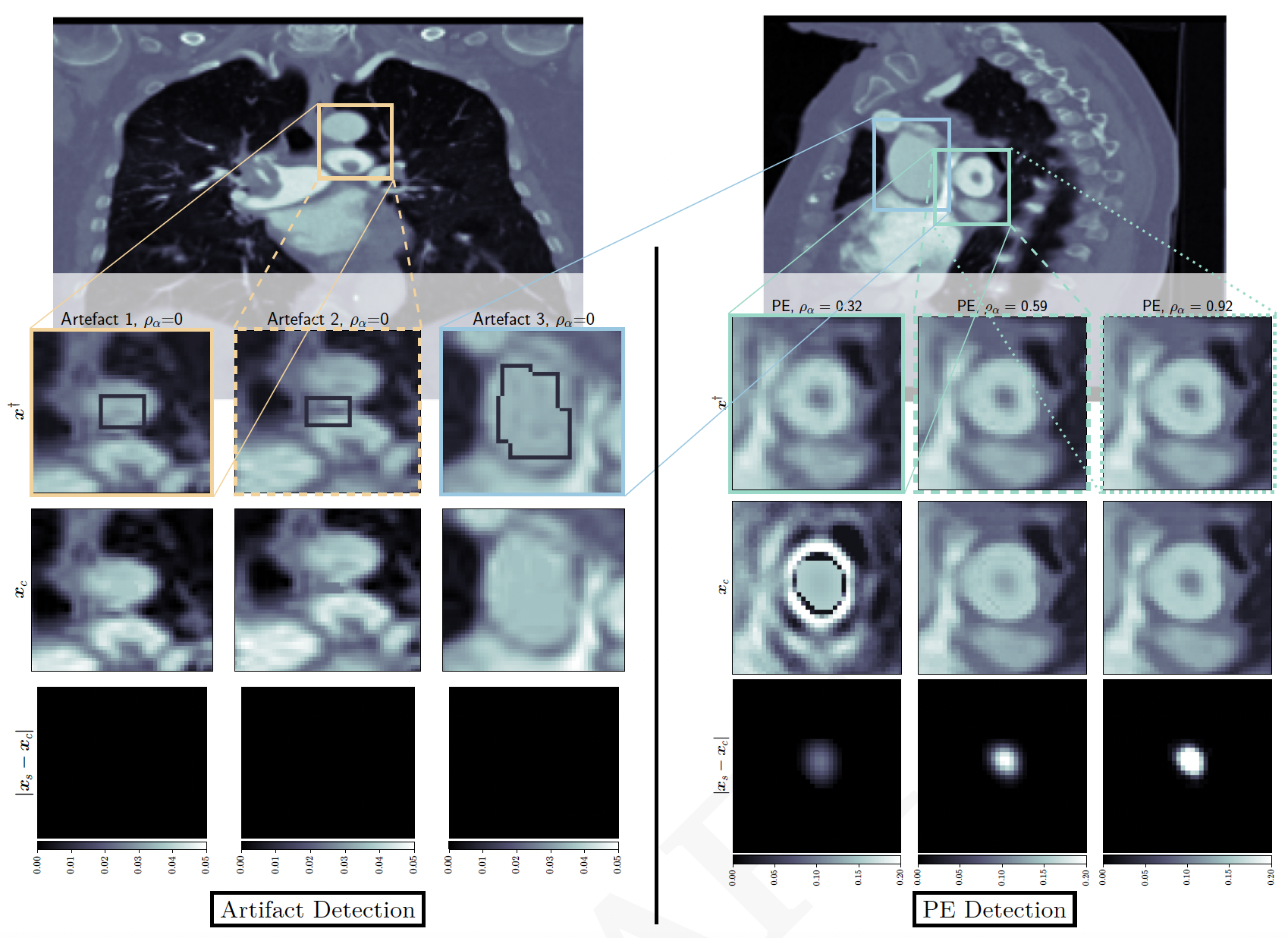}
\end{center}
\caption{
\textbf{Left:} Output of BUQO when used to quantify uncertainty of reconstruction artifacts. The forward problem parameters are chosen to be $(\Nangle,\sigma)=(50,0.175)$ for all the artifacts. 
\textbf{Right:} Output of BUQO when used to quantify uncertainty of PEs, as the value of $\rhoal$ increases. The forward problem parameters are chosen to be (from left to right column): $(\Nangle,\sigma)=(50, 0.007)$, $(\Nangle,\sigma)=(200, 0.035)$ and $(\Nangle, \sigma) = (450, 0.007)$.
\textbf{First row:} MAP estimates, zoomed on the structures of interest.
\textbf{Second row:} Output image $\xc$ from BUQO. 
\textbf{Third row:} Difference images $| \xs - \xc |$.
}\label{fig:rho-qual}
\end{figure*}

\begin{figure*}[t]
\begin{center}

\includegraphics[width=\textwidth]{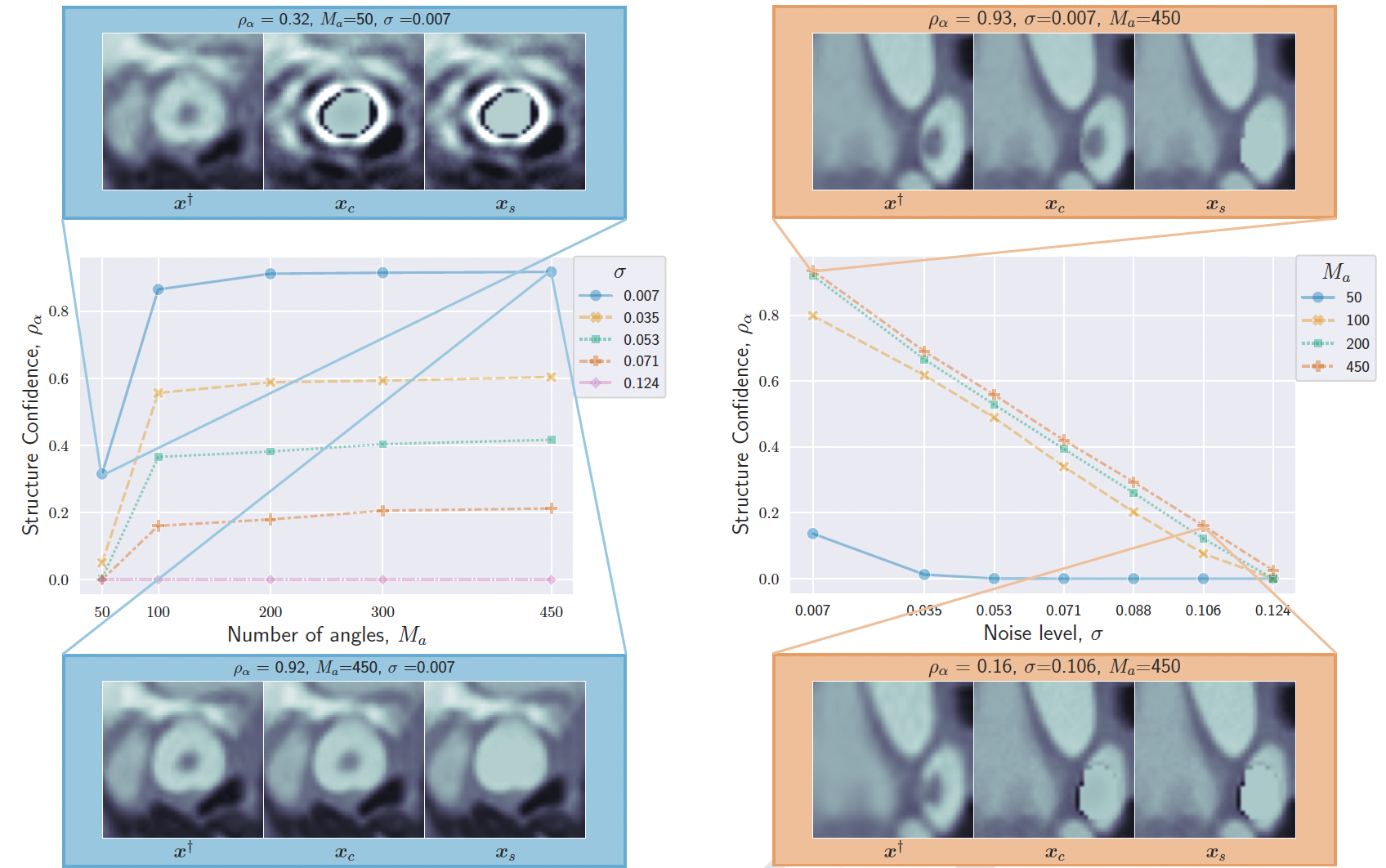}
\end{center}

\caption{Structure confidence $\rhoal$ as a function of number of angles $\Nangle$ ({\bf left}) and noise level $\sigma$ ({\bf right}). High and low structure confidence are illustrated with qualitative examples of $\xc,\xs$ and $\xmap$.
Both plots show that as the data quality (i.e.  number of angles and signal-to-noise ratio) increases, the structure confidence increases too, and we are more certain of the presence of the structure.}\label{fig:pe-detect}
\end{figure*}

\begin{figure}[t]
\centering
\includegraphics[width=8.5cm]{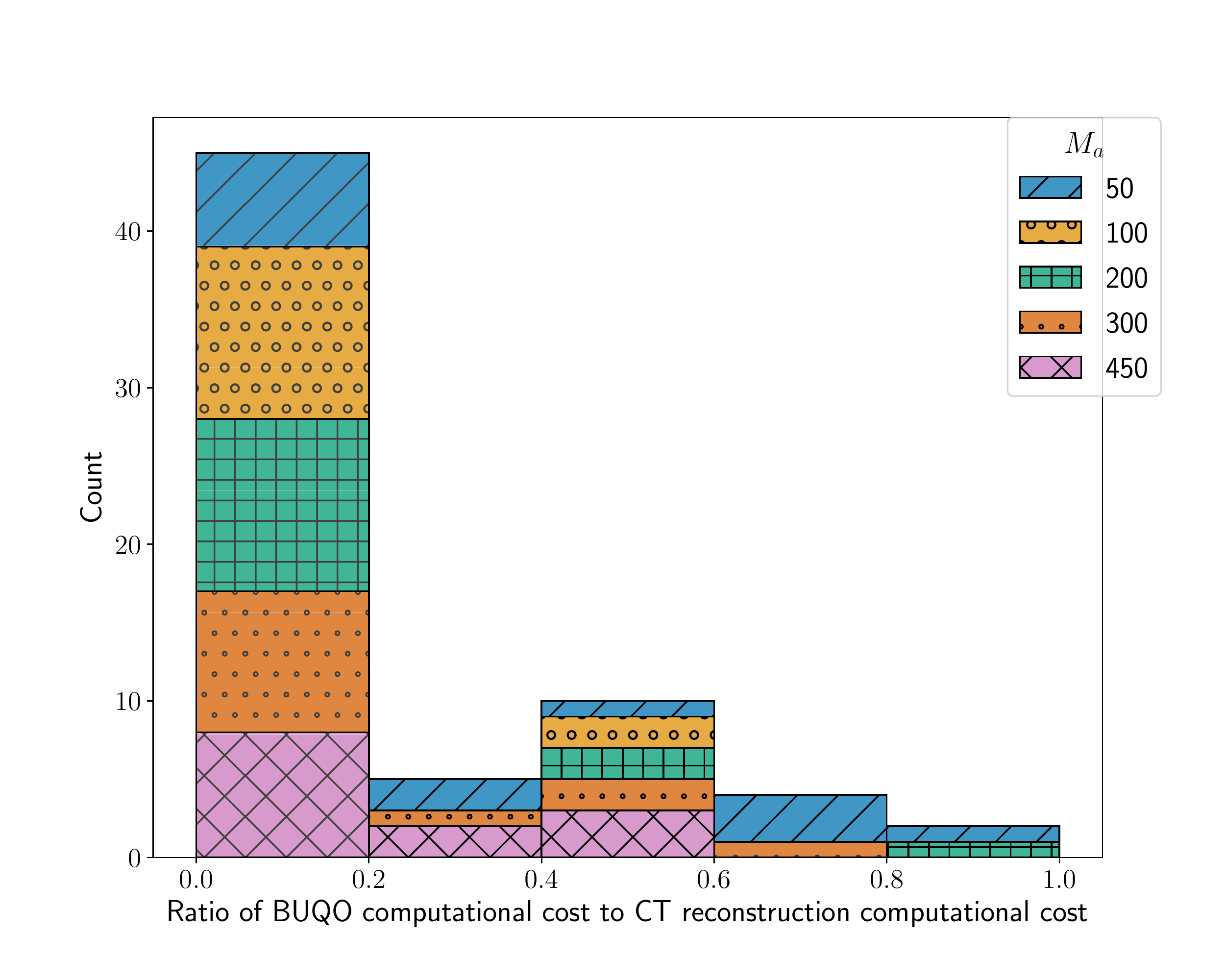}
\caption{The histogram shows the number of forward operator evaluations needed for BUQO convergence as a ratio of the number of forward operator evaluations needed for convergence of the CT reconstruction algorithm. The data is split by the number of angles used in each simulation.}\label{fig:complexity}
\end{figure}

In this section, we present experimental results on synthetic CT data.
We apply the BUQO method to real CT slices that contain a PE and assess the ability of the algorithm to detect the PEs under different noise levels and detector setups. We also apply the BUQO method to test for the presence of reconstruction artifacts that were created when simulating the forward problem. 

\subsection{Experiment Settings}

\subsubsection{Dataset Description}
CTPA was performed on multidetector array scanners, (SOMATOM® Drive and Definition Edge, Siemens Healthineers, Erlangen, Germany). The parameters were as follows; 128 x 0.6 mm slice thickness, 1.2 pitch, 0.5 s rotation time, 145 kVp tube voltage and 120 mAs with automatic dose modulation. 60 mls of non-ionic intravenous contrast medium (iohexol, 350 mg iodine/ml; Omnipaque 350, Amersham Health, England) was administered at 6 ml/s via an 18 G cannula. The acquisition was triggered by bolus tracking of the main pulmonary artery, with a threshold of 100 Hounsfield units (HU) and 4 second delay after triggering.
The study received approval from the Research Ethics Committee and Health Research Authority (IRAS ID 284089). Informed written consent was not required.

\subsubsection{Measurements}
From this data, we consider two slices of reconstructed clinical images containing PEs. Using these slices, we simulate data to study the effect of CT acquisition quality on PE detection. To this end we consider the model described in~\eqref{pb:inv}, with a forward operator $\fwd$ modeling a parallel beam geometry with a fixed number of detectors $\Ndct=450$ and a variable number of acquisition angles $\Nangle \in \{50, 100, 200, 300, 450\}$. We generate $\noise$ in~\eqref{pb:inv} as a realization of an i.i.d. Gaussian noise vector of size $\Nangle \times \Ndct$ and variance $\sigma^2$.
We then reconstruct the CT image by solving equation \eqref{pb:mingen} to obtain the MAP estimate.

\subsubsection{PE definition}

To create the masks related to the operator $\mask$ in \eqref{def:energy} and \eqref{def:smooth}, we used MITK \cite{wolf2004medical}. 
Two types of masks were created by experienced clinical radiologists: Masks identifying the location of real PEs appearing in the CTPA scans; and masks identifying the location of PE-like artifacts appearing in the CTPA scans due to low quality of the acquired data. 
In Figure~\ref{fig:rho-qual} we show, for both slices, the PEs, and the artifacts of interest arising from the reconstruction process.

The set $\Sset$, as defined in Section~\ref{Sec:method}\ref{Ssec:method:hyp}, captures the pixel profile for an artery that does not have a PE. In the definition of $\Sset$, some parameters related to the energy and smoothness constraints must be chosen (see~\eqref{def:energy} and \eqref{def:smooth}, resp.). We propose to choose them automatically, by looking at histograms of pixel intensities and gradients in a neighborhood of the mask. 
Precisely, we sample pixels around the area of interest and compute the histogram of the intensities of the sampled pixel. Then, in~\eqref{def:energy}, $\mupix$ is set to be the median of this histogram, and $\radpix$ is set to be the maximum of the difference between the upper 60$^{\text{th}}$ percentile and the median; and the difference between the median and the lower 60$^{\text{th}}$ percentile. The same is done to compute $\mugrad$ and $\mupix$ in~\eqref{def:smooth}, but with the histogram of sampled gradients instead.

\subsubsection{Result interpretation}

To assess the effect of the acquisition quality (i.e., noise level $\sigma$ and number of angles $\Nangle$) on the ability of our method to detect true structures, we introduce a structure confidence quantity
\begin{equation}
\rhoal = \frac{\| \xch-\xsh \|_2}{\| \xmap-\xmapS \|_2} \in [0,1].
\end{equation}
If $\rhoal=0$, then $\text{dist}(\Sset, \Calpha)=0$ and we can conclude that there exists an image without the observed structure that lies in the credible set $\Calpha$. 
If $\rhoal>0$, then $\text{dist}(\Sset, \Calpha)>0$, and the null hypothesis is rejected.
The closer to one the value of $\rhoal$ is, the more certain we are that the null hypothesis should be rejected, and thus that the structure of interest is present in the true image.  
In practice, numerical errors must be taken into account, and the two above conditions should be relaxed as $\rhoal \le \delta$ and $\rhoal>\delta$, respectively, for some tolerance $\delta$ to be determined by the user.

Note that $\rhoal$ provides additional information than only an accept/reject hypothesis test. It can be interpreted as a percentage of the structure's energy that is confirmed by the data. So when a selected PE-like structure is probed for UQ, $\rhoal$ provides a percentage of the structure's energy that can be trusted.

In Figure \ref{fig-comparison}, we compare our method to traditional hypothesis testing in statistics. It is therefore natural to interpret $\rhoal$ as being equivalent to a p-value in hypothesis testing. However, accepting or rejecting the null hypothesis in our cases does not depend on some hard threshold on $\rhoal$. There are two reasons for this.
Firstly, traditional hypothesis testing is a frequentist method, where one would typically take the output of models at face value. Our method is a Bayesian method, where one is more interested about prior and posterior distribution. As such, $\rhoal$ is telling us the percentage of the structure that can be explained by the data. Setting a threshold on when to accept or reject the null hypothesis should be an application-specific matter.
Secondly, the method we have proposed does not only generate $\rhoal$, but also generates $\xc$ and $\xs$, whose qualitative contribution to the decision to accept or reject the null hypothesis is as important as the quantitative contribution of $\rhoal$. Figure~\ref{fig:rho-qual} shows images $\xc$ and difference images $| \xc - \xs| $, for different detector settings, and therefore different values of $\rhoal$. It can be seen that non-negative values of $\rhoal$ do not necessarily correspond to images that would be considered normal by a radiologist. However, very high values of $\rhoal$ (close to 1) tend to correspond to high fidelity images, which mimic real CT scans very well.

\subsection{Results}

\subsubsection{Confidence with respect to measurements}

We show in Figure~\ref{fig:pe-detect} the behavior of $\rhoal$ for two assessed PE structures, with respect to the noise level $\sigma$ for a fixed number of angles $\Nangle$ (left), and with respect to the number of angles for a fixed noise level (right). 
It can be observed that the ability of the algorithm to confirm the presence of PEs improves with decreasing noise levels and increasing number of angles. 

For the PE structure in Figure~\ref{fig:pe-detect}(left), we provide additional results in Figure~\ref{fig:rho-qual}(right). The images show the results of BUQO when considering $(\sigma, \Nangle) = (50, 0.007) $, $(\sigma, \Nangle) = (200, 0.035) $, and $(\sigma, \Nangle) = (450, 0.007) $. In particular, the last row shows the differences (in absolute values) between $\xs$ and $\xc$. 
This corresponds to the residual PE structure that is probed by BUQO. It can be seen as a 2D map version of quantity $\rhoal$, giving the intensity value per pixel that is validated by the data.
We can see that when the acquisition quality improves (i.e., $\sigma$ decreases and/or $\Nangle$ increases), the intensity value per pixel that is validated by the data increases.

In Figure~\ref{fig:rho-qual}(left) we show results of BUQO for three PE-like structures that are reconstruction artifacts. For these structures, the last row show that the intensity value per pixel that is validated by the data is equal to 0 (i.e., $\rhoal=0$). Hence our method cannot reject $\text{H}_0$, and the data cannot support the existence of the structure.

\subsection{Complexity}

In our experiments (see Figure \ref{fig:complexity}), we found that the numerical complexity of the proposed uncertainty quantification is usually negligible compared to that of the reconstruction algorithm providing the MAP estimate. 
The computational bottleneck is usually the evaluation of the forward operator and its adjoint. The complexity is assessed in terms of total number of iterations (i.e., number of evaluations of the forward operators and their adjoints) to reach convergence of the algorithms used to evaluate the MAP and for BUQO (primal-dual algorithms in both cases). Convergence is assumed to have occurred when all constrained are satisfied, and the estimates are stable, up to a fixed tolerance.

\definecolor{linecol}{RGB}{40, 42, 54}
\definecolor{fillcol}{RGB}{40, 42, 54}

\section{Discussion}
We have introduced an UQ method in CT imaging that can be used to assess PE-like structures observed in CT scans. We have simulated different acquisition environments by varying the number of measurements and the noise level in the forward problem and used the resulting MAP estimate to investigate the behavior of the proposed method to quantify uncertainty of PE-like structures. 
Our method demonstrates diminishing confidence with a decrease in data quality, while correctly identifying reconstruction artifacts produced in simulation using low quality data. In this closing section, we go over the strengths and weaknesses of the proposed method. 

{\bf Manual annotations.}
The proposed method requires 3 inputs, namely the MAP estimate, the mask that isolates the area under investigation and the set $\Sset$, which represents our prior knowledge. \\
Currently, the mask is the result of a time consuming manual segmentation exercise, done by experienced clinical radiologists which can be replaced by an automatic segmentation algorithm based on deep learning methods \cite{soffer2021deep}.\\
The set $\Sset$ is built making use of a constraint defined in the gradient domain of the image (which is unsuitable for artifacts appearing close to a boundary); and is done by manual sampling (which is time consuming). Instead, the set $\Sset$ could be the result of a data-driven method such as generative appearance models  \cite{cootes2001active} 




{\bf Clinical Use.}
Acute PE carries a significant associated morbidity and mortality and thus improvement in the degree of radiologist certainty in positive identification of acute PEs in clinical practice is paramount. It is also important to improve the degree of radiologist certainty in identifying artifacts as such rather than false positive PEs, in order to avoid inappropriate treatment with anticoagulation and unnecessary bleeding risks. Further work is needed to validate the described method in clinical practice.

\bibliographystyle{plain}
\bibliography{pnas-sample}

\subsection*{Author contribution}
{\bf A.M.R.}: Conceptualization, Methodology, Software, Data Curation, Writing-Original Draft Preparation, Investigation, Visualization. 
{\bf H.K.} Data Curation, Resources, Validation. 
{\bf J.R.} Data Curation, Resources.  
{\bf J.S.} Funding Acquisition.
{\bf J.C.L.R.}: Project Administration, Funding Acquisition, Resources, Validation.  
{\bf M.J.E.}: Conceptualization, Methodology,  Writing-Reviewing and Editing, Supervision, Project Administration, Funding Acquisition.
{\bf A.R.}: Conceptualization, Methodology, Software, Writing-Reviewing and Editing, Supervision, Project Administration, Funding Acquisition.

\subsection*{Author declaration}
J.C.L.R makes the following disclosures: Speaker's fees - Sanofi, Consultancy fees - NHSX, Physician services - HeartFlow, Co-founder and share holder - Heart \& Lung Imaging LTD, Part time employee and share holder - RadNet. J.S has received speakers fees, consultancy fees and travel grants from AstraZeneca, Chiesi, MSD and Janssen Pharmaceuticals

\subsection*{Acknowledgements}
MJE acknowledges support from the EPSRC (EP/S026045/1, EP/T026693/1, EP/V026259/1) and the
Leverhulme Trust (ECF-2019-478). AR acknowledges support from the Royal Society of Edinburgh. All authors were supported by the Research Capability Funding of the Royal United Hospital.

\end{document}